\documentclass[11pt]{article}             
\usepackage{osid}                         %
\usepackage{amsmath} 
\usepackage{amsfonts} 
\usepackage{amssymb} 
\usepackage{latexsym} 
\usepackage{pstricks} 
\usepackage{pst-grad} 
\usepackage{pst-node} 
\psset{framearc=0.0,framesep=0.0truecm,nodesep=8pt,unit=0.75truecm, 
      arrowsize=4pt 2,arrowinset=0.4} 
                               
\newcounter{mycnt} 
\def\themycnt{\thesection\arabic{mycnt}} 
\def\mybenv#1{\refstepcounter{mycnt}%
       \vskip 3pt\noindent{\bf \themycnt~#1}:~} 
\def\myeenv{\hfill\rule{1ex}{1ex}\vskip 3pt} 
\def\qed{\hfill$\Box$} 
\def\ed{\end{document}} 
\def\nn{\nonumber \\} 
\def\Id{\text{I\!d}} 
\def\antip{\textsf{S}} 
\def\ip{\star}			
\def\conv{\ast} 
 
\def\openC{\mathbb{C}}

\def\openZ{\mathbb{Z}} 
\def\grpGL{\sf{GL}} 
\def\grpSp{\sf{Sp}} 
\def\grpSL{\sf{SL}} 
\def\grpO{\sf{O}} 
 
\def\grpH{\sf{H}}

\def\algGL{{\mathfrak{gl}}} 
\def\calR{{\cal R}} 
\def\calS{{\cal S}} 
\def\Hom{\text{Hom}} 
\def\End{\text{End}} 
\def\tr{\text{tr}} 
\def\FinVect{\text{\bf FinVect}} 
 
\def\!{\kern -0.15ex} 
\def\nn{\nonumber\\} 
\def\ot{\otimes} 
\def\pleth{\underline{\otimes}} 
\def\ra{\rangle} 
\def\la{\langle} 
\def\({(\!(} 
\def\){)\!)} 
\def\wgt{{\rm wgt}\,} 
\def\RHD{B. Fauser, P.D. Jarvis, R.C.King : \textit{A Hopf algebraic  
approach to group branchings}} 
\def\LHD{B. Fauser, P.D. Jarvis, R.C.King : \textit{A Hopf algebraic approach to group branchings}} 
\title{A Hopf algebraic approach to the theory of group branchings} 
\author{Bertfried Fauser\\{\footnotesize\it Max Planck Institut f\"ur 
        Mathematik in den Naturwissenschaften, Inselstr. 22-26, D-04103  
	Leipzig, Germany; {\tt fauser@mis.mpg.de}}\\[2ex] 
        Peter D. Jarvis\\{\footnotesize\it University of Tasmania, 
	School of Mathematics and Physics, GPO Box 252-21, 7001 Hobart, TAS,  
	Australia; {\tt Peter.Jarvis@utas.edu.au}}\\[2ex] 
	Ronald C. King\\{\footnotesize\it School of Mathematics, University of  
	Southampton, Southampton SO17 1BJ, England;  
	{\tt R.C.King@maths.soton.ac.uk}} } 
 
\begin{document} 
\maketitle 
\begin{abstract} 
We describe a Hopf algebraic approach to the Grothendieck ring of 
representations of subgroups $\grpH_\pi$ of the general linear group 
$\grpGL(n)$ which stabilize a tensor of Young symmetry $\{\pi\}$. It turns 
out that the representation ring of the subgroup can be described as a 
Hopf algebra twist, with a 2-cocycle derived from the Cauchy kernel 
2-cocycle using plethysms. Due to Schur-Weyl duality we also need to employ the 
coproduct of the inner multiplication. A detailed analysis including combinatorial proofs for our results can be found in 
\cite{fauser:jarvis:king:wybourne:2005a}. In this paper we focus on the 
Hopf algebraic treatment, and a more formal approach to representation rings 
and symmetric functions. 
\end{abstract} 
 
\section{Group representation rings} 
 
We are interested in the representation rings of $\grpGL(n)$ and  its subgroups
described as stabilizers of certain elements $T^{\pi}$ of  Young symmetry
$\pi$. A matrix representation $\rho : \grpGL(n)\rightarrow  GL(m)$, $m\ge n$
is polynomial if the entries of $\rho(g)\in\grpGL(m)$  are polynomials in the
entries of $g\in\grpGL(n)$. The character of the representation $\rho$ is the
central function $\chi_\rho : \grpGL(n) \rightarrow \openC$, $\chi_\rho(g) =
\tr(\rho(g))$. Representations form an Abelian semigroup under the direct sum
$V^\lambda\oplus V^\mu$, which is completed to form the Grothendieck group
$\calR^n=\calR_{\grpGL(n)} (\{V^\lambda\},\oplus)$ using virtual
representations $-V^\lambda$ \cite{knutson:1973a}. The tensor product
$V^\lambda\ot V^\mu=  \oplus^\nu C^\nu_{\lambda\mu} V^\nu$ turns this
structure into a  ring $\calR^n =\calR_{\grpGL(n)}(\{V^\lambda\},\oplus,\ot)$.
We proceed  to the inductive limit $\calR_{\grpGL} = \lim_{\leftarrow} R^n$
since finitely generated representation rings develop syzygies while the limit
ring is free. Finite examples thus require establishing these syzygies by
so-called modification rules. 

We follow \cite{tamvakis:2004a}. $\grpGL(n)$ acts by conjugation on the Lie
algebra $\algGL(n)$ of all  $n\times n$-matrices, hence acts on the
invariant ring  $\text{Pol}(\algGL(n))^{\grpGL(n)}$ with integer coefficients
(in Lie  theory real coefficients). Under some topological restrictions one
can  identify the characters $\chi\in\calR$ bijectively with elements in 
$\text{Pol}(\algGL(n))^{\grpGL(n)}$ via the isomorphism $\phi :
\calR_{\grpGL(n)}  \rightarrow \text{Pol}(\algGL(n))^{\grpGL(n)}$ of class 
functions. 

A particular basis of the representation ring is given by equivalence classes of
irreducible representations $V^\lambda$, labelled by integer partitions $\lambda$
(see  below). For each partition label $\lambda$ there exists a Schur map  (Schur
endofunctor on $\FinVect_\openC$, \cite{macdonald:1979a}) mapping the vector space
$V=\openC^n$ the corresponding Schur module $V^\lambda$, a highest weight
$\grpGL(n)$-module in Lie theory. The character of $V^\lambda$ is the Schur
polynomial $s_\lambda$ that is polynomial in the eigenvalues of $g\in\grpGL(n)$.  

Let $\dim V=n$. On any tensor product $W=\ot^p V$ we have the left $\grpGL(n)$
action and right action of the symmetric group $\calS_p$. As bimodule we have 
$W=\ot^p V=\oplus^\lambda V^\lambda \ot S^\lambda$ by Schur-Weyl duality 
\cite{weyl:1930a,halverson:ram:2003a}.  One considers Young symmetrizers
$Y_\lambda= c_\lambda r_\lambda$  where $c_\lambda$ is the column antisymmetrizer
of the tableau of shape  $\lambda$ and $r_\lambda$ is the row symmetrizer. The
$Y_\lambda$ are  idempotents and reduce $W$ into irreducible parts with respect
to  $\grpGL(n)\times \calS_p$. The Schur module   $V^\lambda$ is the image of the
identity morphisms $W\rightarrow W$ defined as  right multiplication by
$Y_\lambda$. This twofold nature of Schur polynomials will cause the remarkable
self duality of the Hopf algebra studied in the next paragraph. 

\section{Symmetric functions} 
 
\subsection{The Hopf algebra of symmetric functions} 
 
An introduction to symmetric functions can be found e.g. in 
\cite{sagan:1991a,kerber:1999a,macdonald:1979a}, the well known Hopf algebra
structure is discussed in 
\cite{geissinger:1977a,zelevinsky:1981b,thibon:1991a,king:1971a}. Here we 
focus on the Hopf algebraic aspects of symmetric functions related to 
representation rings of the $\grpGL$ groups and their Weyl groups 
$\lim_{\leftarrow}\oplus\calS_p$ using the isomorphism between $\grpGL(n)$
and $\calS_p$ representation rings and $\Lambda=\openZ[x_1,x_2,\ldots]^{\calS}
= \lim_{\leftarrow}\oplus^n \openZ[x_1,\ldots,x_n]^{\calS^n}$  of symmetric 
functions in infinitely many variables. 

Schur functions $s_\lambda$, or in Littlewood's bracket notation $\{\lambda\}$,
are indexed by integer partitions $\lambda=(\lambda_1,\ldots,\lambda_k)
=[1^{r_1},\ldots,p^{r_p}]$, where the $\lambda_i$, ordered by magnitude 
$\lambda_i\ge\lambda_{i+1}$,  are called parts, the $r_i$ are multiplicities, and
are conveniently displayed by Ferrers diagrams (also called Young  diagrams).
Schur functions are given by $s_\lambda(x) = \sum_{T \in ST^{\lambda}}
x^{\wgt(T)}$, where the sum is over all tableaux (fillings) $T$ belonging to the
set  $ST^{\lambda}$ of semi-standard tableaux (column strict, row semistrict) of 
shape $\lambda$. Each summand is a monomial in the variables $x_{1}, x_{2}, 
\ldots, x_{n}$ of degree $n=\vert\lambda\vert=\sum\lambda_i$. The module
underlying $\Lambda$ is spanned by $\openZ$-linear combinations of Schur functions
(irreducible representations). To establish the ring structure we introduce the
\textit{outer multiplication} 
\begin{align} 
V^\lambda\ot V^\mu  
&=\oplus^\nu C^\nu_{\lambda\mu} V^\nu 
&&\stackrel{\phi}{\Leftrightarrow}
&  s_\lambda(x) \cdot s_\mu(x)   
&=\sum_{\nu} C_{\lambda\mu}^{\nu} s_\nu(x).  
\end{align}  
Where the nonnegative integer constants $C_{\lambda\mu}^{\nu}$ are the  famous
Littlewood-Richardson coefficients determined e.g. combinatorially. 

Schur functions are important because they encode characters of irreducible 
representations of the $\grpGL(n)$ groups which by Schur's lemma decompose
into isoclasses. The Schur-Hall scalar product encodes this fact letting Schur
functions be orthogonal by definition $\la~\mid~\ra :\Lambda\ot_\openZ
\Lambda \rightarrow \Lambda$, $\la s_{\lambda}\mid s_{\mu}\ra
=\delta_{\lambda,\mu}$. This implies an elementwise identification of the 
module underlying $\Lambda$ with the dual module $\Lambda^\star = 
\Hom(\Lambda,\openZ)$. $\Lambda^\star$ is \emph{a priori} not an algebra!
However, inspection of classical results shows \cite{fauser:jarvis:2003a} that
we can introduce the \textit{same} outer product on the dual $\Lambda^\star$
reflecting the Frobenius reciprocity. Using the Milnor-Moore theorem this
induces a coalgebraic structure on $\Lambda$ fulfilling the axioms of a Hopf
algebra \cite{fauser:jarvis:2003a}. Schur functions $s_\lambda$ have a life as characters of
$\grpGL(n)$-modules $V^\lambda$ and through the Schur-Weyl
duality mentioned previously are also associated with
irreducible representations of $\calS_p$,
a remarkable incidence. With
$f,h,g\in\Lambda$ we define the  outer coproduct $\Delta : \Lambda \rightarrow
\Lambda\ot \Lambda$ as 
\begin{align} 
\la \Delta(f)\vert g\ot h\ra 
&:= \la f\vert g\cdot h\ra 
  = \la f_{(1)}\vert g\ra\la f_{(2)}\vert h\ra  
&&& 
\Delta(s_\lambda)  
&= \sum_{\eta,\xi} C^\lambda_{\eta\xi}s_\eta\ot s_\xi 
\end{align} 
where we have introduced Sweedler indices $\Delta(f)= \sum_{(f)} f_{(1)}\ot
f_{(2)}$ \cite{sweedler:1969a} neatly to  encode the double sum. The unit of
the outer product is $1$, the constant Schur function $s_{(0)}$, given by
injection of the underlying ring  $\openZ\stackrel{\eta}{\rightarrow}\Lambda$.
The coproduct has a counit $\Lambda\stackrel{\epsilon}{\rightarrow}\openZ$
given as  $\epsilon(s_\lambda)=\delta_{\lambda,(0)}$. Product and coproduct
fulfill the homomorphism property 
\begin{align} 
\Delta(f\cdot g)  
&= (fg)_{(1)}\ot (fg)_{(2)}  
 = f_{(1)}\cdot g_{(1)} \ot f_{(2)}\cdot g_{(2)}  
 = \Delta(f)\Delta(g) 
\end{align} 
showing $\Lambda$'s bialgebra structure. The Hopf algebra $\Lambda$ admits an 
antipode $\antip(s_\lambda) = (-)^{\vert\lambda\vert}s_{\lambda^\prime}$ 
where $\vert\lambda\vert=\sum \lambda_i$, and $\lambda'$ is the partition conjugate to $\lambda$
 (that is, the partition associated with the Young diagram specified by $\lambda$ reflected through its diagonal). 
 
We define the adjoint with respect to the Schur-Hall scalar product, called skew
operation, of the outer multiplication
\begin{align} 
\la s_\mu^\perp\cdot s_\lambda \mid s_\pi\ra  
&=\la s_{\lambda/\mu}\mid s_\pi\ra 
 =\la s_\lambda \mid s_\mu\cdot s_\pi\ra \nn 
s_\mu^\perp(x)\cdot s_\nu
&\equiv s_{\nu/\mu}(x)=\sum_{\lambda} C_{\lambda\mu}^{\nu} s_\lambda(x). 
\end{align} 

Through their association with $\calS_p$-modules, Schur functions inherit a
second, {\it inner}, product determined by the product
$\chi^\lambda\,\chi^\mu=\sum_{\nu}\gamma_{\lambda\mu}^\nu\, \chi^\nu$ of
characters of $\calS_p$-modules, where $\lambda$, $\mu$ and $\nu$ are all
partitions of $p$. We denote this product as $\ip : \calS_p \times \calS_p 
\rightarrow \calS_p$ and it can be dualized using the Schur-Hall scalar  product
as $\la \delta(f)\mid g\ot h\ra := \la f\mid g\ip h\ra$,  $\delta(f)= \sum_{[f]}
f_{[1]}\ot f_{[2]}$. We should not be astonished to note that the convolution
$(\ip,\delta)$ is neither Hopf nor even a bialgebra \cite{fauser:jarvis:2003a}.
However, we will need the coproduct $\delta$ in a prominent place! 

\subsection{Plethysm or composition} 

Schur maps (functors) enjoy composition $s_\lambda\pleth s_\mu \equiv
s_\mu[s_\lambda]$ (Note the index  order!) A linear map $f: V \rightarrow U$
determines a linear map  $s_\lambda(f) : s_\lambda(V) \rightarrow
s_\lambda(U)$ being functorial, i.e. $s_\lambda(f_1\circ f_2) = s_\lambda(f_1)
\circ s_\lambda(f_2)$, and $s_\lambda(\Id_V)=\Id_{s_\lambda(V)}$. We use the
symbol $\pleth$ to distinguish plethysms from ordinary tensor products. In
terms of Schur functions the plethysm, or composition, of the Schur function
$s_\mu$ with the Schur function $s_\lambda$, is given by  $s_\mu[s_\lambda](x)
= s_\mu(y) =\sum_{T \in ST^{\mu}} y^{\wgt{(T)}}$ where the entries in each
tableau are now  taken from the set  $\{y_{i}\,|\,i=1,2,\ldots,m\}$ of
monomials of the Schur function  $s_\lambda(x)$. 

\mybenv{Example}   

Consider $s_{(2)}[s_{(1^2)}](x_1,\ldots,x_4)$. Expand $s_{(1^2)}$ as
$s_{(1^2)}(x_1,\ldots,x_4)  = x_{1}x_{2} + x_{1}x_{3} + x_{1}x_{4} +
x_{2}x_{3} + x_{2}x_{4} + x_{3}x_{4} = y_{1} + y_{2} + y_{3} + y_{4} + y_{5} +
y_{6}$ which leads to the expansion of the composition $s_{(2)}[s_{(1^2)}]$
as  
\begin{align} 
s_{(2)}[s_{(1^2)}]&(x_{1},\ldots,x_{4})  =s_{(2)}(y_{1},\ldots,y_{6})  
=s_{(2)}(x_{1}x_{2},\ldots,x_{3}x_{4}) \nn  
&= y_1^2 + \cdots + y_6^2 + y_1y_2+ \cdots +  y_5y_6 \nn    
&= x_1^2x_2^2 + \cdots + x_3^3x_4^2 + x_1^2x_2x_3
   +\cdots + x_4^2x_2x_3   + 3x_1x_2x_3x_4 \nn  
&= s_{(2^2)}(x_{1},\ldots,x_{4})  +s_{(1^4)}(x_{1},\ldots,x_{4})
\end{align} 
The problem of the evaluation of the plethysm is to expand $s_{(2)}(y)$ 
in the Schur function basis $s_\nu(x)$ with $\nu$ a partition of 
$4=\vert\{2\}\vert+\vert\{1^2\}\vert$.
\myeenv 

Plethysm is tied to representation theory: Consider two groups $\grpGL(m)$,
$\grpGL(n)$, with $m>n$. Let a Schur function $\{\lambda\}$ represent the 
character of a $m$-dimensional irrep of $\grpGL(n)$ surjectively embedded 
in the fundamental representation of the group $\grpGL(m)$ whose character 
is $\{1\}$. The branching process $\grpGL(m)\rightarrow\grpGL(n)$ is then 
described by the injective map $\{1\}\rightarrow\{\lambda\}$ which leads 
to the general formula explaining physical origin of the symbol $\pleth$ 
\begin{align} 
\grpGL(m)\rightarrow\grpGL(n)\,: 
&& 
\{\mu\}\rightarrow \{\lambda\} \pleth \{\mu\} 
\end{align}  
The connection with the previous definition of a  plethysm as a composition
comes about because the $\grpGL(n)$ character   $\{\lambda\}$ is nothing other
than the Schur function $s_\lambda(x)$, with  a suitable identification of
$x$. Its dimension $m$, obtained by  setting all $x_i=1$, is just the number
of monomials $y_i$ in  $s_\lambda(x)$, and $s_\mu(y)$ is the corresponding
$\grpGL(m)$ character  $\{\mu\}$.  In terms of modules, let $V^\lambda$ be the
$\grpGL(n)$-module with  character $\{\lambda\}$ and dimension $m$. This
module may be identified  with the defining $\grpGL(m)$-module $V$ on which
$\grpGL(m)$ acts  naturally.  Then the plethysm $\{\lambda\}\pleth\{\mu\}$
arises as the  character of the  $\grpGL(m)$-module $V^\mu=(V^\lambda)^\mu$
viewed as a  $\grpGL(n)$-module. As a result of this interpretation it is
sometimes  convenient to adopt a suggestive exponential notation for plethysms
$\{\lambda\} \pleth \{\mu\}=\{\lambda\}^{\pleth\{\mu\}}$. 

\subsection{Schur function series} 
 
Littlewood \cite{littlewood:1940a} introduced formal power series 
$\Lambda[[t]]$ ($S$-function series) to encode the process of group branching 
\cite{yang:wybourne:1986a}, i.e. passing over to subgroup characters or 
taking `traces' out. Grothendieck introduced $\lambda$-rings, a special formal
power series ring assigned to any commutative ring formalizing the idea of 
taking antisymmetric powers. $\lambda$-rings contain the full information about 
representations \cite{knutson:1973a}. We will use only basic series of 
mutually inverse $M$ and $L$ series $M_t\cdot L_t=1$ defined as 
\begin{align}
M_t(x) &= \prod_{i\ge 1} (1-x_it)^{-1} = \sum \{m\} t^m \nn 
L_t(x) &= \prod_{i\ge 1} (1-x_it) = \sum (-)^m \{1^m\} t^m 
\end{align}
$S$-function series enjoy point wise sum and product. We consider only series
derived from the basic $L$ and $M$-series applying plethysms, e.g:
\begin{align} 
A_t(x)=L_t(x_ix_j)_{(i<j)} 
&=\prod_{i<j}(1-x_ix_jt) 
 =\sum_\alpha (-1)^{\frac{\vert\alpha\vert}{2}}\{\alpha\}t^{\omega_\alpha} 
 =\{1^2\}\pleth L_t\nn 
C_t(x)=L_t(x_ix_j)_{(i \le j)}  
&=\prod_{i\le j}(1-x_ix_jt)  
 =\sum_\gamma(-1)^{\frac{\vert\gamma\vert}{2}}\{\gamma\}t^{\omega_\gamma} 
 =\{2\}\pleth L_t \nonumber
\end{align} 
together with the inverse series $B=A^{-1}$ and $D=C^{-1}$. In general we 
will consider $M_\pi$-series, $M_\pi=\{\pi\}\pleth M$. These series differ 
dramatically in their Schur function content when expanded as $S$-function
series. E.g. the $A$-series ($C$-series) will encode the process of 
extracting all traces w.r.t. an antisymmetric (symmetric) second rank tensor. 

To every Schur function series $\Phi_t$ we assign a linear form $\phi_t$  via
$\phi_t(s_\lambda)= \la \Phi_t\vert s_\lambda\ra$. We are interested in the
sum of Schur functions contained in $\Phi_t$ dropping  $t$ assuming it is
evaluated at $t=1$. 

\section{Morphisms between group representation rings} 
 
We model subgroups $\grpH_\pi$ of $\grpGL(n)$ groups fixing  a certain tensor
$T^{\pi}$ of Young symmetry type $\{\pi\}$. The simplest  examples are given
by stabilizing a vector\footnote{In some basis ${\bf e_i}$.}   $v^i$, a
symmetric rank 2 tensor\footnote{$(.,.)$ and   $[.,.]$ brackets denote
symmetry and antisymmetry of tensors.} $g^{(i,j)}$ or an antisymmetric rank
2 tensor $f^{[i,j]}$. These tensors give rise to the classical subgroup
branchings 
\begin{align} 
\grpGL(n-1)\subset \grpGL(n)&&&   
\{\lambda\}\mapsto\{\lambda/(\{1\}\pleth L)\}=\{\lambda/M^{-1}\} 
\label{skewM}\\ 
\grpO(n)\subset \grpGL(n)&&&   
[\lambda]\mapsto\{\lambda/(\{2\}\pleth L)\}=\{\lambda/C\}  
\label{skewC}\\ 
\grpSp(n)\subset \grpGL(n)&&& 
\la\lambda\ra\mapsto\{\lambda/(\{1^2\}\pleth L)\}=\{\lambda/A\} 
\label{skewA} 
\end{align} 
where $[.]$ and $\la.\ra$ brackets denote $\grpO(n)$ and $\grpSp(n)$ 
characters. Considering the first case, we need to extract from every 
$\grpGL(n)$ tensor all components in the subspace spanned by $v^i$. One finds 
easily 
\begin{align} 
u^i 
&= (u^i-\la u^i\mid v_j\ra v^j) + \la u^i\mid v_j\ra v^j 
 \quad\Leftrightarrow\quad  
 \{1\} \downarrow \{1\}+\{0\} 
\nn 
T^{(i,j)} 
&=(T^{(i,j)}-(T^{(l,j)}v^i+T^{(i,l)v^j})v_l  
 -T^{(l,k)}v_lv_kv^iv^j)  
\nn 
&+(T^{(l,j)}v^i+T^{(i,l)v^j})v_l 
 +T^{(l,k)}v_lv_kv^iv^j 
 \,\Leftrightarrow\, 
 \{2\} \downarrow \{2\}+\{1\}+\{0\} 
\nonumber  
\end{align}
This is systematically done by skewing with the $M^{-1}$-series as shown in 
eqn. (\ref{skewM}). In a similar way the extraction of traces w.r.t. the 
symmetric and antisymmetric rank 2 tensors can be formalized along the same 
way as the classical results (\ref{skewC}) and (\ref{skewA}) show. It is 
noteworthy that this process encodes the Wick theorem of QFT 
\cite{fauser:2001b}. The branchings are bijections since we have $M\cdot 
M^{-1}=1$ and  $(s_\lambda/\mu)/\nu= s_\lambda/(\mu\cdot \nu)$ Which leads to 
$ (\{\lambda\}/\Phi)/\Phi^{-1}   = \{(\lambda/\Phi)/\Phi^{-1}\}  =
\{\lambda/(\Phi\cdot\Phi^{-1})\} =\{\lambda\} $ obtaining the branchings
$\grpGL(n-1)\uparrow\grpGL(n)$,  $\grpO(n)\uparrow\grpGL(n)$ and
$\grpSp(n)\uparrow\grpGL(n)$; all these are  classical results. 

The skew by a Series can be modeled by using the Hopf algebraic framework  $
s_\lambda/\Phi = (\phi\ot \Id)\Delta(s_\lambda)$  and the $/\Phi$ operators
are called \textit{branching operators} for that  reason. All of the $M_\pi =
\{\pi\}\pleth M$ series have an inverse $M^{-1}_\pi=\{\pi\}\pleth M^{-1}$. The
new characters of the subgroups $H_\pi(n)\subset\grpGL(n)$ stabilizing a 
tensor $T^\pi$ of Young symmetry $\{\pi\}$ are denoted by double parentheses 
$\(\lambda\)_\pi$ where the subscript is usually dropped or even be replaced
by the dimension of the irreducible representation space. We find the 
generalization of the classical branchings 
\begin{align} 
\grpH_\pi(n)\subset \grpGL(n)&&&   
\(\lambda\)\mapsto\{\lambda/(\{\pi\}\pleth L)\}=\{\lambda/M_\pi^{-1}\} 
\label{skewHpi}\\ 
\grpGL(n)\supset \grpH_\pi(n)&&&   
\{\lambda\}\mapsto\{\lambda/(\{\pi\}\pleth M)\}=\(\lambda/M_\pi\)  
\label{skewHpiInv} 
\end{align} 
Hence we find isomorphisms of the modules underlying the representation  
spaces
\vskip-16pt
\begin{align}  
\rnode{A}{\calR_{\grpGL}=\calR\Big(\{\lambda\},\oplus\Big)} 
&&&  
\rnode{B}{\calR_{\grpH_\pi}=\calR\Big(\(\lambda\),\oplus\Big)}  
\ncline[offset=3pt]{->}{A}{B} 
\Aput{\{\lambda\}/M^{-1}_\pi} 
\ncline[offset=3pt]{->}{B}{A} 
\Aput{\(\lambda\)/M_\pi} 
\end{align} 
\vskip6pt 
For labelling problems of many particle states in quantum mechanics this 
process allows to compute states and energy levels. However, one would  like
to reobtain the ring structure of the subgroup representation rings 
$\calR_{\grpH_\pi}=\calR\Big(\(\lambda\)_\pi,\oplus,\ot_\pi\Big)$, which is 
considered to be a hard problem. Due to the restrictions imposed by 
extracting traces the tensor product of two reduced, say  $\grpO(n)$,
representations is in general no longer in an obvious way  a direct sum of
$\grpO(n)$ representations.  Also the Weyl group changes
\cite{halverson:ram:2003a} and the Schur-Weyl  duality needs a different Weyl
group, e.g. the hyperoctahedral group in case of $\grpO(n)$. 

\section{Hopf algebra twists of group representation rings} 
 
\subsection{Twists of representation rings} 
 
The main aim of the present paper is to explain how Hopf algebra  twists can
be used to establish the desired product formulae for subgroup  characters. 

To do this, we introduce some Hopf algebra cohomology along classical lines
\cite{sweedler:1968a,brouder:fauser:frabetti:oeckl:2002a}. We define
$n$-cochains $c_n : \Lambda^n \rightarrow \openZ$, multilinear forms of 
$n$-arguments. The linear forms $\phi$ associated to the branching operator 
$/\Phi$ is hence a 1-cochain. Having a Hopf algebra $\Lambda$, we can define  
on all endormorphisms $f,g : \Lambda\rightarrow \Lambda$ the convolution 
product $\conv : \End \Lambda\times \End\Lambda \rightarrow \End \Lambda$ 
$(f\conv g)(x)= f(x_{(1)})g_(x_{(2)})$ this product can be generalized
in a straight forward manner to $n$-cochains with multiplication taken in $\openZ$. 

We define a multiplicatively written coboundary operator $\partial_n$ mapping 
$n$-cochains $c_n$ to $(n+1)$-cochains $c_{n+1}$ as 
\begin{align} 
c_{n+1}=\partial^i c_n(x_0,\ldots,x_n)&= 
\left\{ 
\begin{array}{cl} 
\epsilon(x_0)c_n(x_1,\ldots,x_n) & i=0 \\ 
c_n(x_1,\ldots,x_ix_{i+1},\ldots,x_n) & i\in\{1,\ldots,n-1\} \\ 
c_n(x_1,\ldots,x_{n-1})\epsilon(x_n) & i=n 
\end{array} 
\right.  
\end{align}
which can be used to
give $\partial_n = \partial^0c_n \conv \partial^2c_n^{-1} \conv 
\ldots \conv \partial^n_nc_n^{\pm1}$ having alternating signs. If
$\partial_nc_n=\epsilon^{n+1}$, $c_n$ is closed, if $c_{n+1}=\partial_nc_n$
$c_{n+1}$ is exact. Cohomology neatly classifies the linear 1-cochains
associated to Schur function series. Group like series having closed
associated 1-chains, e.g. $M$, $L$, $V$, etc. give rise to homomorphisms
with respect to the branching, havinging an unaltered product structure. 
\mybenv{Theorem \cite{fauser:jarvis:2003a}} 
Let $G$ be a group like Schur function series, i.e.  $\Delta(G)=G\ot G$.
The associated 1-chochain is closed $\partial_1  g=\epsilon^2$. The associated
representation ring $\calR_{\grpGL\downarrow 
H_G}\Big(\(\lambda\)_{G},\oplus,\ot\Big)$ has the same tensor product and
is homomorphic to the representation ring of $\grpGL(n)$.  
\myeenv
An example is the $\grpGL(n)\downarrow\grpGL(n-1)$ branching and its  inverse.
More subtle situations are obtained with 2-cocycles which are exact  hence
derivable from 1-cochains. The 2-cocycle condition assures that the new 
twisted product remains associative. Let $(\partial_1\phi)(x,y) 
=\phi^{-1}(x_{(1)})$ $\phi^{-1}(y_{(1)}) \phi(x_{(2)}\cdot y_{(2)})$ 
One introduces the twisted product w.r.t. this 2-cocycle. We give 
two equivalent formulae  
\begin{align} 
\(\lambda\)_\phi \cdot_\phi \(\mu\)_\phi  
&=  \Big(\(\lambda\)_\phi/\Phi\, \cdot\, \(\mu\)_\phi/\Phi\Big)/\Phi^{-1} \\ 
\(\lambda\)_\phi \cdot_\phi \(\mu\)_\phi  
&= (\partial\phi)\Big(\{\lambda\}_{(1)},\{\mu\}_{(2)}\Big)  
    \Big(\kern-0.5ex\Big(\{\lambda\}_{(2)} \cdot \{\mu\}_{(2)}
    \Big)\kern-0.5ex\Big)_\phi 
\end{align} 
These twists induce not graded but only filtered multiplications due to the
traces which are extracted from the original characters. We state our main 
result about formal characters restricting ourselves to $S$-function series 
$M_\pi$.  
\mybenv{[$\pi$-Newell-Littlewood] Theorem 
\cite{fauser:jarvis:king:wybourne:2005a}} 
\label{piNewellLittlewood} 
The representation ring $\calR_{\grpH_\pi}$ of a subgroup $\grpH_\pi$ 
of $\grpGL(n)$ stabilizing a tensor $T^{\pi}$ of Schur symmetry $\{\pi\}$  
is obtained by a Hopf algebra twist of the character ring  
$\calR_{\grpGL}\Big(\{\lambda\},\oplus,\ot\Big)$ of the $\grpGL(n)$ with respect 
to the subgroup characters $\(\lambda\)_\pi$ and the product deformation 
induced by the Schur function series $M_\pi=\{\pi\}\pleth M$ resp. its 
associated 1-cochain $m_\pi$ ($\otimes_{m_\pi}\equiv\,\cdot_{m_\pi}$) 
\begin{align} 
\calR_{\grpGL}\Big(\{\lambda\},\oplus,\ot\Big) 
&\downarrow  
\calR_{\grpH_\pi}\Big(\(\lambda\)_\pi,\oplus,\ot_{m_\pi}\Big)  
\end{align}  
These branchings are ring isomorphisms, hence the $\uparrow$ direction is 
obtained from $M^{-1}_\pi$. 
\myeenv 

\subsection{Technical details} 
 
Passing to examples needs concrete combinatorial formulae. The key result is the
outer coproduct of $M_\pi$ series. While 
\cite{fauser:jarvis:king:wybourne:2005a} provides formulae for coproducts of 
the plethysm of a Schur polynomial by a Schur function series, we stick to a
\mybenv{Corollary 
\cite{fauser:jarvis:king:wybourne:2005a}} 
\label{corollary} 
For any partition $\pi$, the coproduct of the series 
$M_\pi=\{\pi\}\pleth M(x)\equiv M^{\{\pi\}}$ reads 
\begin{align} 
\label{prop1} 
\Delta M_\pi(x)  
&= (M_\pi)_{(1)}\otimes (M_\pi)_{(2)} = M_\pi(x,y) \nn   
&= (M\otimes M)^{\Delta\{\pi\}}(x,y) 
= M^{\{\pi\}_{(1)}}\otimes M^{\{\pi\}_{(2)}}(x,y)  \nn 
&= M_\pi(x)M_\pi(y)\sum_{\sigma_1\ldots,\sigma_k}\prod_{\xi,\eta<\pi} 
\prod_{l=1}^{C^\pi_{\xi\eta}} 
s_{\{\xi\}\pleth\{\sigma_k\}}(x) 
s_{\{\eta\}\pleth\{\sigma_k\}}(y) 
\end{align} 
where the $C^\pi_{\xi\eta}$ are the Littlewood-Richardson coefficients 
of outer multiplication. The summations over the $\sigma_j$ are 
formally over all Schur functions. The proper cut part of the coproduct is 
defined as $K(x,y)=(M_\pi)_{(1^\prime)}\ot (M_\pi)_{(2^\prime)} = 
\sum\prod\prod s_{\xi\pleth\sigma}(x)s_{\eta\pleth\sigma}(y)$.  
\myeenv 
This remarkable and formidable formula hides somehow a peculiarity. Namely 
that it contains inner coproducts stemming from the Weyl group $\calS_p$. This 
can be seen from a Lemma which was needed to prove the corollary 
\mybenv{Lemma \cite{fauser:jarvis:king:wybourne:2005a}} 
Let $\delta(x)=x_{[1]}\ot x_{[2]}$, $\Delta(x)=x_{(1)}\ot x_{(2)}$, then
\newline  
$\Delta(\{\mu\}\pleth \{\lambda\})  
 =\,\{\mu\}\pleth(\Delta(\{\lambda\})) 
 =\,(\{\mu\}_{[1]}\pleth\{\lambda\}_{(1)})\ot 
    (\{\mu\}_{[2]}\pleth\{\lambda\}_{(2)})$ 
\myeenv 
This generalizes to Corollary (\ref{corollary}) by right  distributivity of
the plethysm. The interplay between Weyl group and  structure of the twist on
the representation ring employs the famous  Cauchy kernel ${\sf
C}(x,y)=\sum_\xi s_\xi(x)s_\xi(y)$, which enters  the twists in a complicated
fashion. It is the \textit{inner coproduct}  of the $M$ series $\delta
M=M_{[1]}\ot M_{[2]}={\sf C}(x,y)$ and the  coefficients $\{m\}$ of the $M$
series are the units of inner products  of irreps of $\calS_m$. This result
allows an easy 
\newline 
{\bf Proof of Theorem \ref{piNewellLittlewood}:} 
We make use of duality, and of the fact that the proper cut 
part of the coproduct of $M_\pi$ has an inverse to calculate the product 
of  $\grpH_\pi(n)$ characters directly in terms of Schur functions: 
\begin{align} 
\la \(\mu\)_\pi&\cdot \(\nu\)_\pi \mid s_\rho \ra 
 = 
\la \mu\otimes \nu \mid L_\pi \ot L_\pi \cdot \Delta\,s_\rho \ra \nn 
&=\la \mu\ot \nu \mid 
(M_{\pi^\prime_{(1)}}L_{\pi^\prime_{(1)}})\otimes 
(M_{\pi^\prime_{(2)}}L_{\pi^\prime_{(2)}}) 
L_\pi\ot L_\pi \cdot \Delta\,s_\rho \ra \nn 
&=\la \mu\otimes \nu \mid 
M_{\pi^\prime_{(1)}}\otimes M_{\pi^\prime_{(2)}} \,\cdot\, 
L_{\pi^\prime_{(1)}}\otimes L_{\pi^\prime_{(2)}} \,\cdot\, 
L_\pi\otimes L_\pi\cdot\Delta\,s_\rho \ra \nn 
&= 
\la \mu/M_{\pi^\prime_{(1)}}\ot \nu/M_{\pi^\prime_{(2)}} \mid 
\Delta L_\pi \cdot \Delta s_\rho \ra \nn 
&= 
\la \mu/M_{\pi^\prime_{(1)}}\ot \nu/LM_{\pi^\prime_{(2)}} \mid 
\Delta (L_\pi \cdot s_\rho) \ra \nn 
&= 
\la  
\(\mu/M_{\pi^\prime_{(1)}}\,\cdot\, 
        \nu/M_{\pi^\prime_{(2)}}\)_\pi\mid 
s_\rho \ra 
\end{align} 
The conclusion follows from nondegeneracy of the Schur scalar product, and 
completeness of the Schur basis. 
\qed 
 
\section{Examples} 
 
This section shall give a look-and-feel idea of what kind of groups may be
expected and  how their character theory looks like. 

\subsection{Representation rings of classical groups} 
 
Special cases of the general formula are the classical results. For the
branchings w.r.t. $L=M^{-1}_{\{1\}}$,  $A=M^{-1}_{\{1^2\}}$ and
$C=M^{-1}_{\{2\}}$. The $M$ series is group like inducing no twist providing
representation rings intertwined by a change of basis (irreducibles) and
identity maps on direct sum and tensor product. 

Remarkably the symplectic and orthogonal cases lead to the same deformation! 
All $M_\pi$ coproducts are of the form
$\Delta(M_\pi)(x)=M_\pi(x)M_\pi(y)$ $K(x,y)$ where $K(x,y)$ is a complicated 
expression obtained from proper cuts of the coproduct, i.e. such parts which 
do not contain an identity component and Cauchy kernels. The proper cuts of 
$\{2\}$ and $\{1^2\}$ are identical! $\Delta^\prime(\{2\}) = 
\Delta^\prime(\{1^2\}) = \{1\}\ot\{1\}$ The theorem \ref{piNewellLittlewood} 
simplifies to 
\mybenv{Newell-Littlewood Theorem} 
\begin{align} 
\grpO(n)  &&  
[\lambda]\cdot_c[\mu] &= \sum_\xi [\lambda/\xi\,\cdot\,\mu/\xi] \\ 
\grpSp(n) &&  
\la\lambda\ra\cdot_a\la\mu\ra &= \sum_\xi \la\lambda/\xi\,\cdot\,\mu/\xi\ra 
\end{align} 
\vskip -3ex 
\myeenv 
Having used formal characters, finite dimensional examples have to cope with
syzygies reinduced by so-called modification rules. Classically one has only
two cases dealt with by case-by-case studies \cite{king:1971a}. The general
theory, having infinitely many cases, needs a not yet available formalism. 

\subsection{Nonclassical groups} 
 
If we fix in $\grpGL(n)$ an epsilon tensor of Young symmetry $\{1^n\}$  unique
up to a multiplicative constant, then we find $\grpH_{1^n} =  \grpSL(n)$
providing a systematic treatment of $\grpSL(n)$ groups. However the 
representation ring $\calR_{H_{1^n}}$ is \textit{not} that of the 
$\grpSL(n)$ groups in the inductive limit.  

Consider $\grpH_{1^3}(4)\subset \grpGL(4)$ choosing a basis for 
$T^{1^3}=\eta$ given as 
\begin{align} 
\eta_{pqr} 
&= 
\left\{ 
\begin{array}{cc} 
 \epsilon_{abc} & a\wedge b \wedge c\in{1,2,3} \\ 
 0 & \text{else} 
\end{array}\right. 
\end{align} 
The subgroup $\grpH_{1^3}$ is characterized by $A^x_p A^y_q A^z_q 
\eta_{pqr} = \eta_{xyz}$ which splits into parts containing an index $4$ or 
not. This gives a $3+1$ decomposition of the matrices into blocks leading to 
\begin{align} 
\grpH_{1^3}\,\ni\,[M]  
&= \left( 
\begin{array}{cc} 
B_{3\times 3} & D_{3\times 1} \\ 
0_{1\times 3} & C_{1\times 1} 
\end{array} 
\right) 
\end{align}
where $0$ is the $1\times 3$ zero-matrix, $C\not=0$ and $\det(B)=1$.   This is
an affine algebra, $D$ playing the role of translations.  In general, the
groups $\grpH_\pi$ are semisimple and may not be  reductive or can be
discrete. 

The interpretation of formal characters of $\grpH_{1^3}(4)$ needs modification rules,
required for all $\(\mu\)$ with $\mu$ of length $\ell(\mu)=4$. In our example
this can only arise in those cases for which $\lambda$ also has length $4$. Since
$\{1^4\}=\varepsilon$ is the  character of the $1$-dimensional determinant
representation of $\grpGL(4)$ it follows that $ 
\{\lambda_1,\lambda_2,\lambda_3,\lambda_4\}=\varepsilon^{\lambda_4}\ 
\{\lambda_1-\lambda_4,\lambda_2-\lambda_4,\lambda_3-\lambda_4,0\}$.
Applying this to the $\grpH_{1^3}$ characters $\(\lambda\)_{\dim}$ gives: 
\begin{align} 
\(1111\)_{-3} &=\{1^4\}_{1}-\(1\)_{4} 
               =\varepsilon\(0\)_1-\(1\)_{4} \nn 
\(2111\)_{-12}&=\{2111\}_{4}-\(2\)_{10}-\(11\)_{6} 
               =\varepsilon\(1\)_{4}-\(2\)_{10}-\(11\)_{6} \nn 
\(2211\)_{-17}&=\{2211\}_{6}-\(21\)_{20}-\(111\)_{3} \nn 
              &=\varepsilon\(11\)_{6}-\(21\)_{20}-\(111\)_{3} \nn 
\(2221\)_{-8} &=\{2221\}_{4}-\(211\)_{11}-\(1111\)_{-3}-\(1\)_{4} \nn 
              &=\varepsilon\(111\)_{3}-\(211\)_{11} \nn 
\(2222\)_{3}  &=\{2222\}_1-((2111\)_{-12}-\(2\)_{10} \nn 
              &=\varepsilon^2\(0\)_1-\varepsilon\(1\)_{4}+\(11\)_{6} \\ 
\(3111\)_{-30}&=\{3111\}_{10}-\(3\)_{20}-\(21\)_{20} 
               =\varepsilon\(2\)_{10}-\(3\)_{20}-\(21\)_{20} \nonumber
\end{align} 
This provides a collection of modification rules to be applied in the case 
$\(\mu\)$ with $\ell(\mu)=4$. A complete set of modification rules, 
including those appropriate to $\(\mu\)$ with $\ell(\mu)>4$ should be
established.  
 
Character tables and further details may be found in 
\cite{fauser:jarvis:king:wybourne:2005a}. Examples for products of subgroup
characters for $\grpH_{1^3}(4)$ are: 
\begin{align}
\begin{array}{c|l} 
\cdot & \(2\)_{10} \\ 
\hline\hline 
\(2\)_{10}  & \(4\)_{35}+\(31\)_{45}+\(22\)_{30} \\  
\(11\)_{6}  & \(31\)_{45}+\(211\)_{11}+{\red\(1\)_{4}} \\ 
\(3\)_{20}  & \(5\)_{56}+\(41\)_{84}+\(32\)_{60} \\ 
\(21\)_{20} & \(41\)_{84}+\(32\)_{60}+\(311\)_{26}+\(221\)_{14} 
       +\(2\)_{10}+\(11\)_{6} \\ 
\(111\)_{3} & \(311\)_{26}+{\green\(2111\)_{-12}} 
       +{\red\(2\)_{10}+\(11\)_{6}} \\  
\end{array} 
\nonumber 
\end{align}
\small{ 

\begin{thebibliography}{99} 
 
\bibitem{brouder:fauser:frabetti:oeckl:2002a} 
Ch.~Brouder, B.~Fauser, A.~Frabetti, and R.~Oeckl. 
\newblock {Quantum field theory and Hopf algebra cohomology [formerly: Let's
twist again]}. 
\newblock {\em J. Phys. A: Math. Gen:} 37:5895--5927, 2004. 
\newblock hep-th/0311253. 
 
\bibitem{fauser:2001b} 
B.~Fauser. 
\newblock On the {H}opf-algebraic origin of {W}ick normal-ordering. 
\newblock {\em Journal of Physics A: Mathematical and General}, 34:105--115, 
  2001. 
\newblock hep-th/0007032. 
 
\bibitem{fauser:jarvis:2003a} 
B.~Fauser and P.~D.~Jarvis. 
\newblock {A Hopf laboratory for symmetric functions}. 
\newblock {\em J. Phys. A: Math. Gen:}, 37(5):1633--1663, 2004. 
\newblock math-ph/0308043. 
 
\bibitem{fauser:jarvis:king:wybourne:2005a} 
B.~Fauser, P.~D.~Jarvis, R.~C.~King, and B.~G.~Wybourne. 
\newblock {New branching rules induced by plethysm}. 
\newblock {\em math-ph/0505037}, pages 1--40, 2005. 
 
\bibitem{geissinger:1977a} 
L.~Geissinger. 
\newblock {Hopf algebras of symmetric functions and class functions}. 
\newblock pages 168--181, 1977. 
\newblock Springer Lecture Notes, 579. 

\bibitem{halverson:ram:2003a}
T.~Halverson, and A.~Ram.
\newblock{\em Partition algebras}.
\newblock{preprint 2003, to appear in European J. of Combinatorics}
 
\bibitem{kerber:1999a} 
A.~Kerber. 
\newblock{Applied finite group actions} 
\newblock{Springer-Verlag, Berlin 1999} 
 
\bibitem{king:1971a} 
R.~C.~King. 
\newblock {Modification rules and products of irreducible representations of 
  the unitary, orthogonal, and symplectic groups}. 
\newblock {\em J. Math. Phys.}, 12(8):1588--1598, 1971. 
 
\bibitem{knutson:1973a} 
D.~Knutson. 
\newblock {\em {$\lambda$-Rings and the Representation Theory of the Symmetric 
  Group}}. 
\newblock Springer-Verlag, Berlin, 1973. 
\newblock Lecture Notes in Mathematics 308. 
 
\bibitem{littlewood:1940a} 
D.~E.~Littlewood. 
\newblock {\em {The Theory of Group Characters}}. 
\newblock Oxford University Press, Oxford, 1940. 
 
\bibitem{macdonald:1979a} 
I.~G.~Macdonald. 
\newblock {\em {Symmetric functions and Hall polynomials}}. 
\newblock Calderon Press, Oxford, 1979. 
\newblock [2nd edition 1995]. 
 
\bibitem{sagan:1991a} 
B.~E.~Sagan. 
\newblock {\em {The Symmetric Group: Representations, Combinatorial Algorithms, 
  and Symmetric Functions}}. 
\newblock Springer-Verlag, New-York, 1991. 
\newblock [2nd revised printing 2001]. 
 
\bibitem{sweedler:1968a} 
M.~E.~Sweedler. 
\newblock {Cohomology of algebras over Hopf algebras}. 
\newblock {\em Trans. Am. Math. Soc.}, 133:205--239, 1968. 
 
\bibitem{sweedler:1969a} 
M.~E.~Sweedler. 
\newblock {\em {Hopf Algebras}}. 
\newblock W. A. Benjamin, INC., New York, 1969. 

\bibitem{tamvakis:2004a}
H.~Tamvakis.
\newblock{The connection between representation theory and Schubert 
 calculus}.
\newblock{\em L'Enseignement Mathematique}, 50:267--286, 2004.
 
\bibitem{thibon:1991a} 
J.-Y.~Thibon. 
\newblock {Hopf algebras of symmetric functions and tensor products of 
  symmetric group representations}. 
\newblock {\em International Journal of Algebra and Computation}, 
  1(2):2007--221, 1991. 
 
\bibitem{weyl:1930a} 
H.~Weyl. 
\newblock {\em {The classical groups, their invariants and representations}}. 
\newblock Princeton University Press, Princeton, N.J., 1930 
  [1946. 2d ed., with supplement]. 
 
\bibitem{yang:wybourne:1986a} 
M.~Yang and B.~G. Wybourne. 
\newblock {New $S$ function series and non-compact Lie groups}. 
\newblock {\em J. Phys. A: Math. Gen.}, 19:3513--3525, 1986. 
 
\bibitem{zelevinsky:1981b} 
A.~V.~Zelevinsky. 
\newblock {\em {Representations of Finite Classical Groups: A Hopf Algebra 
  Approach}}. 
\newblock Springer Verlag, Berlin, Heidelberg New York, 1981. 
\newblock LNM 869. 
 
\end{thebibliography}
%
 
} 
\end{document}